\documentclass[twocolumn]{aa}
\usepackage[utf8]{inputenc}
\usepackage{natbib}
\usepackage{graphicx}
\usepackage{placeins}
\usepackage{color}
\usepackage{amsmath}
\usepackage{natbib}
\usepackage{arydshln}
\usepackage{tikz}
\usepackage{hyperref}
\usepackage{orcidlink}
\newcommand{\FIG}[1]{#1}
\usepackage{xcolor}
\usepackage{soul}

\title{Coronal rain formation in a two-fluid approximation}
\author{B. Popescu Braileanu
          \inst{1,2}
          \and
          R. Keppens \inst{1}\orcidlink{0000-0003-3544-2733}}
\titlerunning{Two-fluid coronal rain}
\authorrunning{Popescu Braileanu and Keppens}
\institute{Centre for mathematical Plasma Astrophysics, KU Leuven, 3001 Leuven, Belgium \and
Department of Physics and Technology, University of Bergen, Norway,\\ \email{beatrice.braileanu@uib.no} }

\date{}

\date{Received XXXX; Accepted XXXX}

\abstract{
{\it Context.}
Coronal rain results from thermal instability in the solar corona, where runaway in-situ cooling causes plasma to condense and drain along the magnetic lines. 
Coronal rain is observed in 3D spine-fan magnetic configurations.
The reconnection of the magnetic field lines around the null point creates jets, seen as denser structures traveling along the field lines. 
As these dense regions evolve,  thermal instability can set in and ultimately form coronal rain.
\\{\it Aims.}
In this paper we study the importance of  partial ionization effects in the formation of  coronal rain in the late evolution of 3D spine-fan magnetic reconnection in the solar corona.
\\{\it Methods.}
We use a two-fluid model consisting of neutral and charged particles coupled by collisions, where ionization recombination processes are taken into account. 
The 3D magnetic field configuration consists of two spine-fan structures and associated null points. We use localized resistivity around the nulls to induce reconnection by driving the bottom 
boundary below one of the two null points. 
To trigger the thermal instability, we here investigate how magnetic reconnection generates flows that lead to the accumulation of higher-density structures along magnetic field lines. 
\\{\it Results.}
The  dynamics associated with the spine-fan magnetic reconnection produces current sheets around the null point and flows along the field lines.
Blobs similar to coronal rain  start to appear after 400 seconds in the simulation domain which follow the field lines from the direction of the perturbed null point. 
  The temperature drop is accompanied by recombination of  charged particles.
    This leads to a drastic increase in the neutral density contrast within the rain blobs, while charges demonstrate the typical two orders of magnitude increase in density as found in single fluid settings. 
\\{\it Conclusions.}
Recombination effects become important in coronal rain evolution when the temperature drops considerably in the condensed structures.
The neutrals are slowed down by recombination, producing decoupling in velocity at the size of the blob, but inside the condensing structure,
the neutrals can move faster across the field lines, creating small scale structures. 
The incomplete elastic collisional coupling produces decoupling in velocity, and associated more efficient frictional heating for the neutrals,
followed by decoupling in temperature.
This study presents a novel two-fluid approach to coronal rain, showing that incorporating two-fluid effects is essential for accurately capturing its dynamics.
}

\begin{document}

\maketitle

\section{Introduction}
\label{sec:intro}
Coronal rain is primarily observed in post-flare loops and active regions of the solar corona and plays a crucial role in understanding coronal heating, plasma dynamics, and mass transport within the solar atmosphere \citep{2018je,2020AHeinzel}.
Chromospheric evaporation  initiates hot plasma inflow during a solar flare, while coronal rain marks the cooling and condensation outflow, completing the mass-energy cycle
\citep{sahin2024}.
Single fluid magnetohydrodynamic (MHD) simulations of post-flare rain only recently progressed from 2D \citep{Ruan_2021} to 3D settings \citep{Ruan2024}, while more quiescent rain in loops and arcades started in multi-dimensional MHD in chromosphere to coronal settings \citep{Fang2013,Fang2015,Moschou2015,Xia2017,xiaohong_cr}. Active region rain was demonstrated recently to be linked to thermal instability in self-consistent radiative MHD, with sub-photosphere to coronal evolutions \citep{Lu2024}.
Observations showed that coronal rain is a multithermal phenomenon, appearing as clumped structures organized in strands \citep{antolin}. The density contrast across the blob interface is around 100 \citep{scullion2016}. 
The coronal rain blobs have width of 300-700~km and a broad velocity distribution, peaking below 50~km/h  \citep{antolin2023}.
Larger velocities of the falling blobs, up to 50–100~km/s, have also been observed \citep{Mason_2019}.
Two-fluid simulation of the evolution of coronal rain blobs showed large decoupling in velocity due to the elastic collisions between ions and neutrals and associated frictional heating \citep{david}.

The coronal rain blobs appear as the nonlinear outcome of the thermal instability caused by radiative cooling of the plasma. The thermal instability was introduced conceptually by \cite{parker} and studied theoretically in the linear MHD assumption in the seminal paper by \cite{field}. 
Local coronal volume simulations confirmed the thermal instability as the trigger for the formation of condensations in the MHD models with optically thin radiative cooling curves corresponding to the solar corona \citep{niels,ti1,ti2,veronika}.
As the rain blobs represent a typically hundredfold increase in density and decrease in temperature, it is clear that plasma-neutral effects can become important as the thermal instability proceeds. None of the in-situ formation studies on coronal rain addressed this aspect in detail, although a recent 1D setup initiated this in fixed loop settings \citep{veronika2025}.

Analytical studies in the two-fluid plasma-neutral approach showed that the presence of neutrals decreases the linear growth rate of the thermal instability when there is a larger fraction of neutrals or when they are coupled more strongly to the charges \citep{ti2f}. The decrease in temperature inside the condensed regions triggers recombination of the charges, and the fraction of neutrals increases so much so that the contrast in the neutral density is orders of magnitude larger than that in the charged density \citep{ti2f}. 
Similar important ionization/recombination processes have been identified in reconnection scenarios, or in the presence of shocks. Indeed, they stabilize the current sheets and decrease the growth of the coalescence instability in the nonlinear phase \citep{Murtas_2022}. The inclusion of inelastic processes and the ionization potential in a two-fluid approximation  can lead to a significant departure from MHD results in slow shock formation \citep{ionrecBen}. In this paper, we will perform 3D two-fluid simulations where reconnection and thermal instability interplay, as relevant for coronal rain.

Coronal rain was observed to be formed as a consequence of the impulsive heating following reconnection events in the solar corona \citep{crObs1} and it is frequently observed 
in postflare loops \citep{Mason_2019}. 
Unlike 2D reconnection,  which happens only at ``X''-type null points, 3D reconnection happens in regions of  electric field concentrations, at null points, 
but also at separators and quasi-separators \citep{li-priest,Pontin2022}.   
The reconnection rates in a 3D setup are usually smaller than in 2.5D and they converge in the limit of weak reconnection with strong guide field \citep{Daldorff_2022}.

Null points are ubiquitous in the solar corona, where small patches of magnetic field emerges through the photosphere, having opposite polarity than the 
main surrounding field. In 3D the magnetic field around null points include isolated field lines, called ``spines'' and a surface of ``fan'' field lines, both structures 
starting or ending near the null point \citep{li-priest,pontin3}. The 3D null points have been studied in a linear field topology approximation around the null point, where 
the magnetic field vector is proportional to the position vector measured from the null point \citep{pontin1,priest1}. 
The simplest null point in this approximation, usually considered in the analytical studies, is the ``proper radial null point'', 
where the spine is perpendicular to the fan and the field lines in the fan are radial, a particular case of the 
general linear null point when the spine is not perpendicular to the fan and the field lines in the fan can be curved  \citep{li-priest,priest1,pontin1,pontin3}.

Small incompressible shearing flows perturbing the spine of the radial null point creates pure fan reconnection \citep{pontin3}, 
but generally, the shearing driving produces spine-fan reconnection \citep{priest1}, which is the most common 3D null point reconnection \citep{li-priest}.
The Lorentz force tends to increase the displacement created by the driving, producing a collapse of the spine towards the fan. As a result, the current sheet formed 
near the null point is inclined with respect to the fan plane with an angle which increases with increasing driving velocity, 
therefore the current sheet contains parts of both spine and fan \citep{pontin1}. 
The current sheet, formed in the vicinity of the null point, becomes more localized and more intense while the null point collapses \citep{priest1}.

In many cases the spine is an open field line, and this situation usually occurs in coronal hole regions \citep{Mason_2019}. 
The jets produced by reconnection in this spine-fan configuration have sizes varying from less than 10~Mm  to hundreds of Mm,
and condensations have been observed for mostly all the intermediate size jets of $\approx100$~Mm \citep{Mason_2019}. The smaller scale condensations (of sizes less than 10~Mm) 
are difficult to observe due to the instrument resolution and the inherent line-of-sight ambiguity for determining both velocity and size \citep{Mason_2019}.
Simulations using a 3D spine-fan magnetic reconnection setup provide a realistic representation of solar magnetic activity. 
Velocity driving of the bottom boundary in these simulations produces an accumulation of current density around the null point \citep{rec1,rec2,Pontin_2013,nitin}.
The magnetic field reconnects in the region around the null point, leading to intense localized energy release. Plasma flows develop along the field lines, guided by the spine and fan structures, transporting energy and momentum away from the reconnection site \citep{Pontin_2013}.  
Observations showed slippage along flare ribbons following flare events moving in opposite directions,   
with velocity usually around 20-40~km/s, but also reaching 400-450~km/s \citep{li-priest}.
The reconnection rate, quantified by the parallel electric field integrated along the field lines in the diffusion region around the null point, has a peak value for the field lines situated in the fan plane \citep{rec2}.
The current density accumulated near the null point is sensitive to the driving \citep{rec2}
The reconnection rate time evolution follows the pattern
of the current density evolution, both decrease when the driving stops, and the system relaxes \citep{pontin2,Pontin_2013}.

In this paper we study the formation of coronal rain  during the late evolution of spine-fan 3D magnetic reconnection. We use a two-fluid approach with ionization recombination processes taken into account. Section \ref{sec:setup} describes the numerical setup, the results are presented in Section \ref{sec:results} and conclusions in Section \ref{sec:concl}.

\section{Numerical setup}
\label{sec:setup}

The 3D computational domain corresponds to a coronal only region of size $-6.8$~Mm~$\le x \le 3$~Mm, $-3$~Mm~$\le y \le 3$~Mm, and  $0 \le z \le 3$~Mm and
 is covered by a grid with base resolution of $144\times80\times64$ and 3 levels of refinement. Therefore, the effective resolution is 
$576\times320\times256$ points and cell sizes: $dx=17.014$~km, $dy=18.75$~km, and $dz=11.719$~km. Hence, we effectively resolve structure down to about few tens of km, compared to the observationally inferred blob sizes of order few hundred km \citep{obsCR1,antolin2023,scullion2016}.  

\begin{figure*}
\FIG{
\includegraphics[width=16cm]{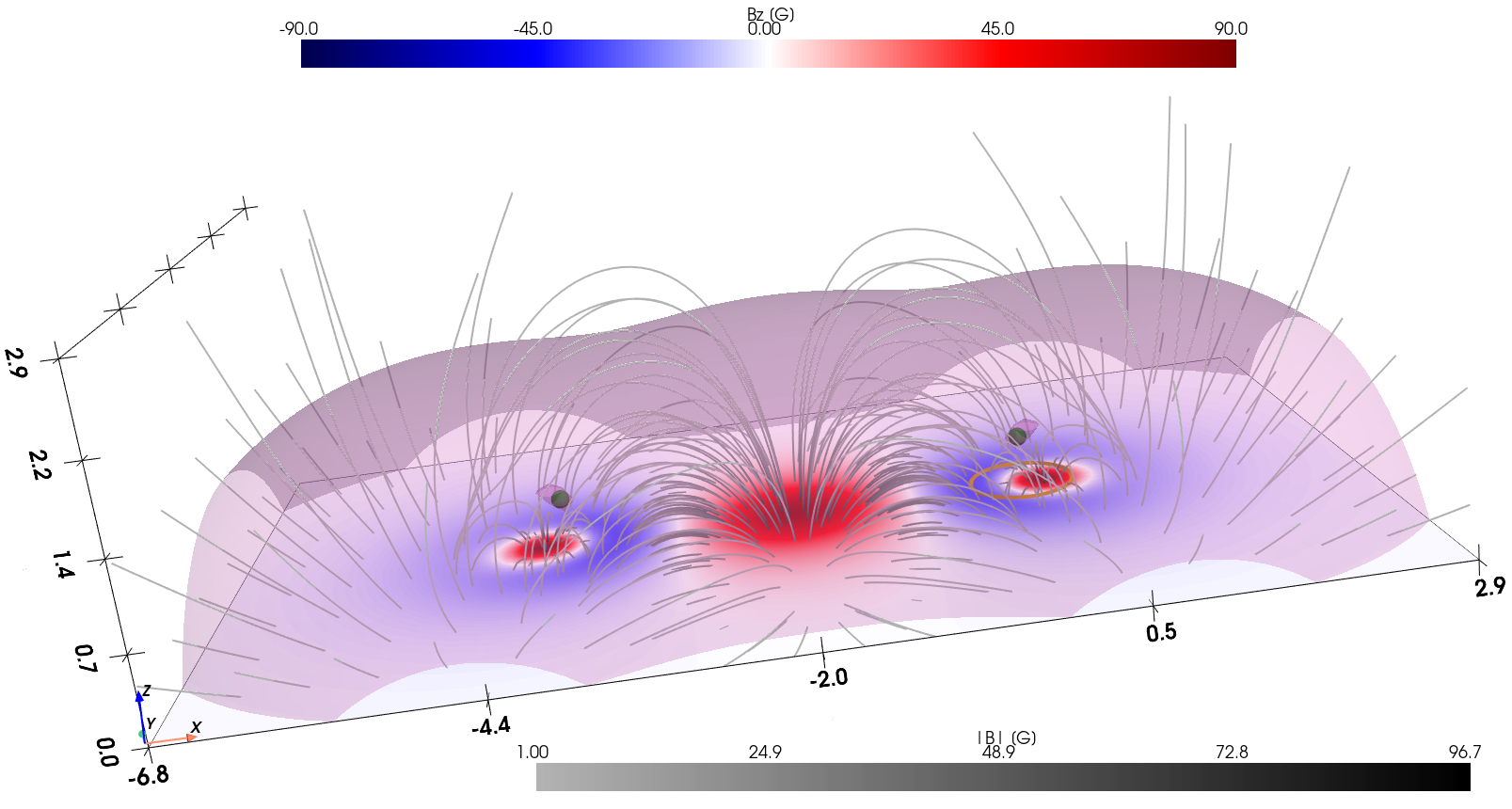}
}
\caption{Initial configuration of the magnetic field. The vertical component of the magnetic field is plotted at the bottom of the domain. 
The two null points are indicated by small spheres. The field lines are plotted using a gray colormap, which shows the magnitude of the magnetic field. The orange circle below one of the null points indicates the location of the velocity perturbation defined by Eq.~\ref{eq:vdr}, having the center at the projection of the  perturbed null point on the bottom boundary and the radius  equal to $2 \sqrt{{\sigma_v}/{2}} = 0.447$~Mm. The isocontours of $\beta_{\rm plasma}=1$ are shown by a semi-opaque violet mesh, showing two regions, one encapsulating the two null points and an outer region.}
\label{fig:bini}
\end{figure*}

The spine-fan magnetic field configuration is very similar to the setup used by \cite{Pontin_2013}. The magnetic field is a superposition of five magnetic point charges  situated below the bottom boundary, resulting in a divergence-free potential magnetic field:
\begin{eqnarray}
\label{eq:bfield}
N_p=5\,,\nonumber\\
x_p=\{-4.1, -3.9,   -1.9,     0.1,    0.3 \}\,,\nonumber\\
y_p = \{  0.0,  0.0,    0.0,     0.0,    0.0 \}\,,\nonumber\\
z_p = \{ -0.4, -1.0,   -1.0,    -1.0,   -0.4 \}\,,\nonumber\\
b_p=\{ 0.3, -1.0,    1.0,    -1.0,    0.3\}\,,\nonumber\\
{B_0}_x(x,y,z) = B_{\rm 00} \sum_{j=1}^{N_p} {b_p}_j \frac{x-{x_p}_j}{r_j^3}\,,\nonumber\\
{B_0}_y(x,y,z) = B_{\rm 00} \sum_{j=1}^{N_p} {b_p}_j \frac{y-{y_p}_j}{r_j^3}\,,\nonumber\\
{B_0}_z(x,y,z) = B_{\rm 00} \sum_{j=1}^{N_p} {b_p}_j \frac{z-{z_p}_j}{r_j^3}\,,\nonumber\\
\text{where }r_j=\sqrt{(x-{x_p}_j)^2 + (y-{y_p}_j)^2 + (z-{z_p}_j)^2}\,.
\end{eqnarray}
We used the value $B_{\rm 00}=100~G$ so we have a typical active region field at the bottom of our coronal box.
The magnetic field configuration is symmetric with respect to $x=-1.9$, with  two magnetic null points
located at: $(x_{\rm N1}, y_{\rm N},z_{\rm N})$ and $(x_{\rm N2}, y_{\rm N},z_{\rm N})$, where
$x_{\rm N1}= 0.123\,,y_{\rm N}=0\,,z_{\rm N}=0.42, x_{\rm N2}=-3.923$. The configuration of the magnetic field is shown in Figure~\ref{fig:bini}, with the two null points indicated by two small spheres. 
These nulls are preferential sites for reconnection events, and we will enhance this possibility by adopting an anomalous, spatially localized resistivity.

The resistivity localized around the two null points  has  a profile very similar to that used in \cite{jets}:
\begin{eqnarray}
    \label{eq:etaLoc}
\eta(x,y,z)=(\eta_0-\eta_1) \Big[  \nonumber\\
 \text{exp}\left(-\frac{(x-x_{\rm N1})^2 + (y-y_N)^2 + (z-z_N)^2} {2 \sigma_r}\right)  +\nonumber \\ 
\text{exp}\left(-\frac{(x-x_{\rm N2})^2 + (y-y_N)^2 + (z-z_N)^2} {2 \sigma_r} \right) \Big] + \eta_1\,,
\end{eqnarray}
where $\eta_0 = 8~\Omega m$, $\eta_1 = 0.8~\Omega m$, and $\sigma_r=0.2$~Mm.

We use open boundary conditions in the $x$ and $y$ directions and for the top boundary in the $z$-direction.
For the bottom $z$-boundary we use 
a driver during the whole period of the simulation for the $x$-component of the velocity, centered around the projection of the first null point:
\begin{equation}
\label{eq:vdr}
V_x=V_0 \text{sin}^2\left( \frac{2  \pi  t}{P}\right)  \text{exp}\left(-\frac{(x-x_{\rm N1})^2 + (y-y_N)^2}{\sigma_v}\right)
\end{equation}
where $V_0$=10~km/s, P=100~s and $\sigma_v$=0.1~Mm. 
The location of the driver is indicated by an orange circle in Figure~\ref{fig:bini}.
This driving is adopted for both neutrals and charges. 
The $y$ and $z$ components of the velocities on the bottom boundary are antisymmetric, which implies that they vanish on the boundary and all the other variables are symmetric. 
This bottom boundary condition is similar to that used by \cite{Pontin_2013}. The simulation is run for 598~s.

We solve numerically the nonlinear two-fluid equations, also shown in \cite{paper2f}, which in this case (without gravity and Hall term, but including the radiative losses) are: 
\begin{eqnarray}
\label{eqs:2fl_start}
\frac{\partial \rho_{\rm n}}{\partial t} + \nabla \cdot \left(\rho_{\rm n}\mathbf{v}_{\rm n}\right) = 
S_{\rm n}\,,
 \\
\frac{\partial \rho_{\rm c}}{\partial t} + \nabla \cdot \left(\rho_{\rm c}\mathbf{v}_{\rm c}\right) = 
-S_{\rm n} \,,
\\
\frac{\partial (\rho_{\rm n}\mathbf{v_{\rm n}})}{\partial t} + \nabla \cdot \left(\rho_{\rm n}\mathbf{v_{\rm n}} \mathbf{v_{\rm n}}  + p_{\rm n} \mathbb{I}  \right)
 = \mathbf{R}_{\rm n} \,,
\\
\frac{\partial (\rho_{\rm c}\mathbf{v_{\rm c}})}{\partial t} + \nabla \cdot \left[\rho_{\rm c}\mathbf{v}_{\rm c} \mathbf{v}_{\rm c} +  
\left( p_{\rm c} + \frac{1}{2}B^2\right) \mathbb{I} -\mathbf{B}\mathbf{B} \right] 
 =  -\mathbf{R}_{\rm n} \,,
\\
\frac{\partial e^{\rm tot}_{\rm n}}{\partial t}  +  \nabla \cdot \left[ \mathbf{v}_{\rm n} \left(e^{\rm tot}_{\rm n} + p_{\rm n}\right)    \right ]  
 =  M_{\rm n} \,,
\\[0.15cm]
\frac{\partial e^{\rm tot}_{\rm c}}{\partial t}  +  \nabla \cdot \Big[ \mathbf{v}_{\rm c} \left( 
e^{\rm tot}_{\rm c} +  p_{\rm c} +\frac{1}{2} B^2 \right) - \mathbf{B} (\mathbf{v}_{\rm c} \cdot \mathbf{B}) 
 \Big ]   
\nonumber\\[0.12cm]
  = \eta J^2  
-M_{\rm n} -\rho_c^2 \Lambda + \rho_{\rm c0}^2 \Lambda_0 \,,
\\
\frac{\partial \mathbf{B}}{\partial t} +  \nabla  \cdot \left( \mathbf{v}_{\rm c} \mathbf{B}  - 
\mathbf{B} \mathbf{v}_{\rm c} \right)= \eta \mathbf{J} .
\label{eqs:2fl_end}
\end{eqnarray}
The above set of Eqs.~(\ref{eqs:2fl_start})-(\ref{eqs:2fl_end}) is written for conserved variables: mass densities, momentum and total energies:
\begin{equation}
e^{\rm tot}_{\rm c} = e_{\rm c}+\frac{1}{2}\rho_{\rm c} v_{\rm c}^2 +\frac{1}{2} B^2\,;\quad
e^{\rm tot}_{\rm n} = e_{\rm n}+\frac{1}{2}\rho_{\rm n} v_{\rm n}^2  \,.
\end{equation}
The pressure is prescribed using the ideal equation of state,
\begin{equation}
p_i = (\gamma-1)e_i\,,\text{for $i$=n,c}\,.
\end{equation}
The collisional terms are:
\begin{eqnarray}
S_{\rm n} = \rho_{\rm c} \Gamma^{\rm rec} - \rho_{\rm n}\Gamma^{\rm ion}\,,\nonumber\\
\mathbf{R}_{\rm n} = \rho_{\rm c} \mathbf{v}_{\rm c} \Gamma^{\rm rec}  - \rho_{\rm n} \mathbf{v}_{\rm n} \Gamma^{\rm ion} + 
\rho_{\rm n} \rho_{\rm c} \alpha (\mathbf{v}_{\rm c} - \mathbf{v}_{\rm n})\,,\nonumber\\
M_{\rm n} = \frac{1}{2} \rho_{\rm c} v_{\rm c}^2 \Gamma^{\rm rec}  
- \frac{1}{2}\rho_{\rm n} v_{\rm n}^2 \Gamma^{\rm ion} 
+  \frac{1}{2} ({v_{\rm c}}^2 - {v_{\rm n}}^2) \rho_{\rm n} \rho_{\rm c} \alpha
\nonumber\\
+ \frac{1}{\gamma-1} 
\left ( \rho_{\rm c} T_{\rm c} \Gamma^{\rm rec} - \rho_{\rm n} T_{\rm n} \Gamma^{\rm ion} \right) 
\nonumber\\
+\frac{1}{\gamma-1} (T_{\rm c} - T_{\rm n})\rho_{\rm n} \rho_{\rm c} \alpha\,.
\end{eqnarray}
where the elastic collisional parameter,
\begin{equation} \label{eq:alpha_el} 
\alpha = \frac{2}{{m_H}^{3/2} \sqrt{\pi}} \sqrt{ k_B T_{cn}}  \Sigma_{in} \,,
\end{equation}
with $T_{cn}= (T_c +T_n)/2$
and the ionization/recombination rates,
\begin{equation} \label{eq:gamma_rec}
\Gamma^{\rm rec}  \approx \frac{n_e}{\sqrt{T_e^*}}      2.6 \cdot 10^{-19} \,\,\,\, {\rm s^{-1}}
,\end{equation}
\begin{equation}  \label{eq:gamma_ion}
\Gamma^{\rm ion}  \approx n_e A \frac{1}{X + \phi_{\rm ion}/{T_e^*}}\left(\frac{\phi_{\rm ion}}{T_e^*}\right)^K  e^{-\phi_{\rm ion}/T_e^*} \,\,\,\,  {\rm s^{-1}}
,\end{equation}
where 
$\phi_{\rm ion} = 13.6 eV$, 
$T_e^*$ is electron temperature in eV, and constants have values
$A = 2.91 \cdot 10^{-14}$, 
$K$ = 0.39, and $X$ = 0.232 in the SI unit system.
The radiative cooling term $-\rho_c^2 \Lambda$ is compensated by a background heating: $\rho_{\rm c0}^2 \Lambda_0$ which cancels the cooling exactly at time $t=0$, 
the simulation starting from  thermal equilibrium.
For numerical reasons, the variables are split into time-dependent and time-independent quantities, the 
equivalent equations solved by the code being Eqs.~(2.7)-(2.14) from \cite{ti2f}.

We use initially uniform thermodynamic background plasma parameters, characteristic to the solar corona, as used  
in \cite{ti2f}: the coronal plasma pressure is set to the constant $p_0=0.032$~Pa, while the plasma density is given the constant value $\rho_0=2.3\times 10^{-12}$~kg/m$^3$. This makes the coronal temperature at a value of around $10^6$~K.  Since we start with a solar coronal volume, the ionization fraction corresponding to ionization/recombination equilibrium is very high, having an almost negligible initial fraction of neutrals: $1-\xi_i=1.019\times10^{-6}$. Therefore we use the same initial conditions for thermodynamic variables and the same equations as in the simulation presented in Section 4a) in  \cite{ti2f}, but in a 3D setup and a different configuration of the force-free background magnetic field. As the magnetic field varies in 3D, the magnitude of the field defined in Eq.~\ref{eq:bfield} has the maximum value of 99.55~G  on the bottom boundary at the points (-4.1112,0,0) and (0.3112,0,0) and the minimum value of 0 at the two null points, resulting in an
initial range of plasma beta from $8.1\times10^{-4}$ to $\infty$. The isocontours of $\beta_{\rm plasma}=1$ are shown in Figure~1 by a semi-opaque isosurface revealing an inner region around the null points of about 200~km size and an outer region, further away from the magnetic point sources. 

The cooling curve used for the optically thin radiative losses is ``Colgan\_DM''.
The  equations are solved numerically with the code {\tt MPI-AMRVAC} \citep{amrvac,paper2f}.
We used a finite volume approach for spatial discretization, employing an HLL Riemann solver \citep{Toro1997} with a ``koren'' limiter for variable reconstruction. We used the semi-implicit IMEX\_ARS3 scheme for the time integration. The refinement criterium was based on the charged fluid density. 
\section{Results}
\label{sec:results}
\begin{figure}
\FIG{
\includegraphics[width=10cm]{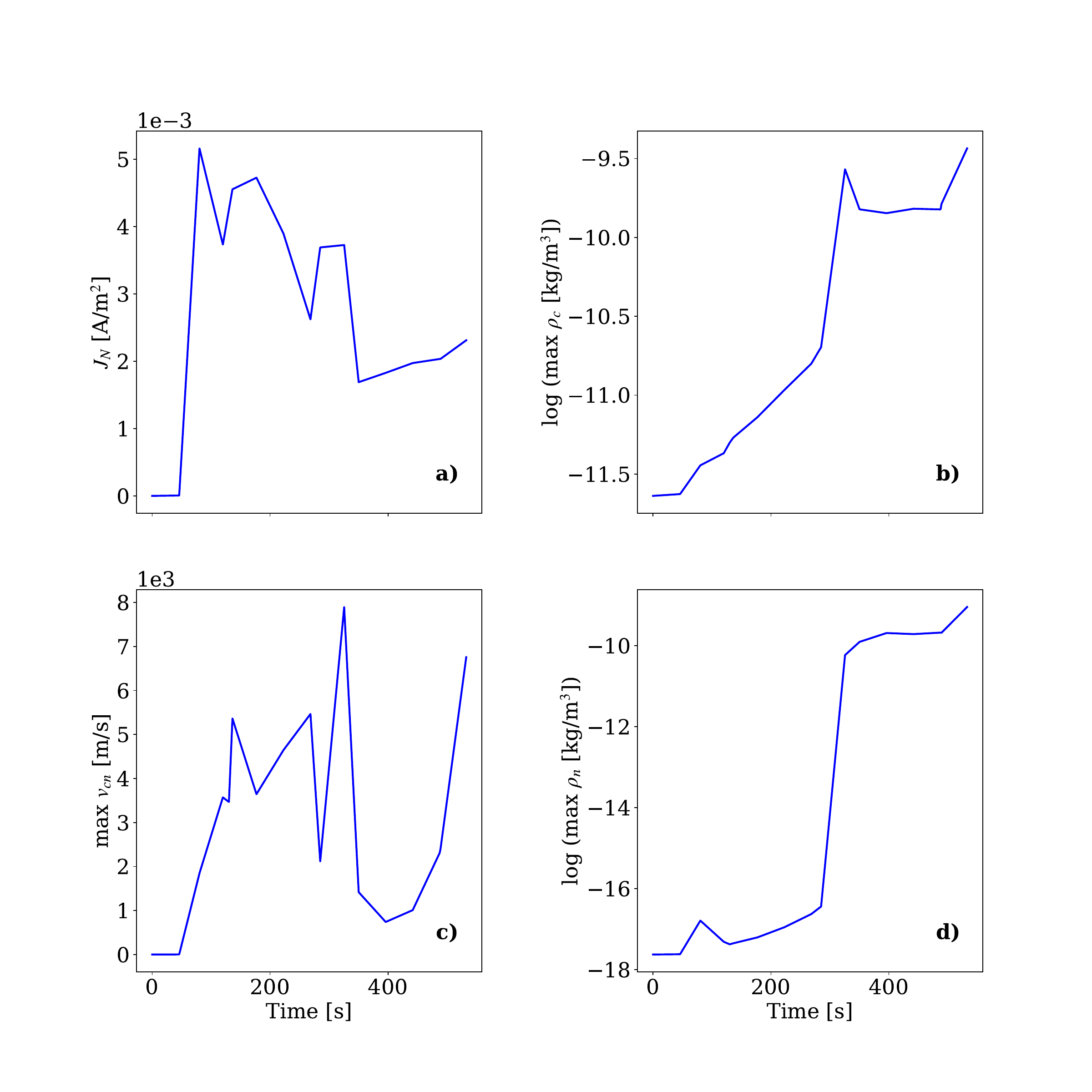}
}
\caption{Time evolution of: \textbf{a)} current density at the null point;
\textbf{b)} logarithmic value of the maximum density  of charges;
 \textbf{c)} maximum value of the magnitude of the decoupling in velocity;
 \textbf{d)} logarithmic value of the maximum density of neutrals. }
\label{fig:tev}
\end{figure}
\begin{figure*}
\FIG{
\includegraphics[width=8cm]{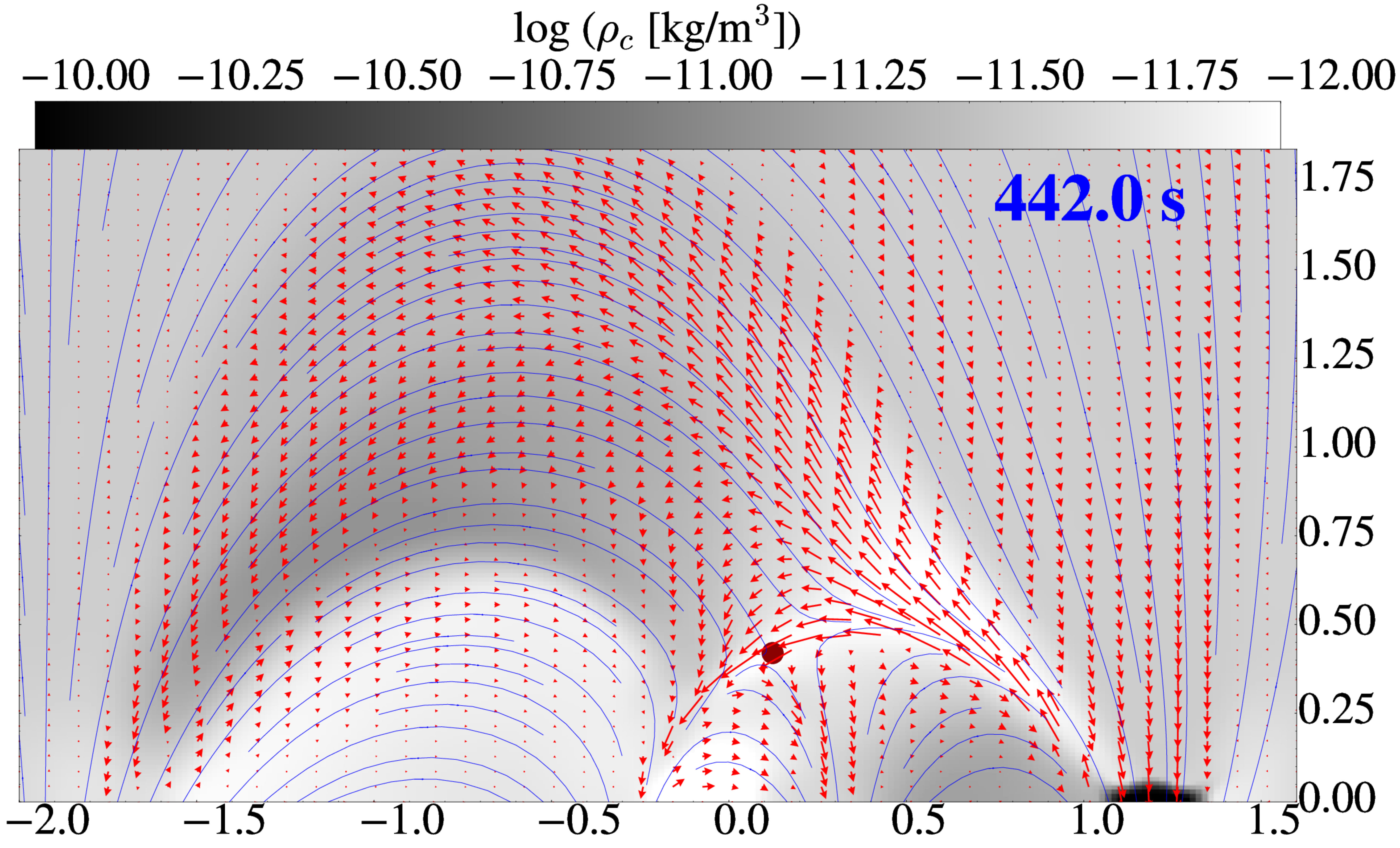}
\includegraphics[width=8cm]{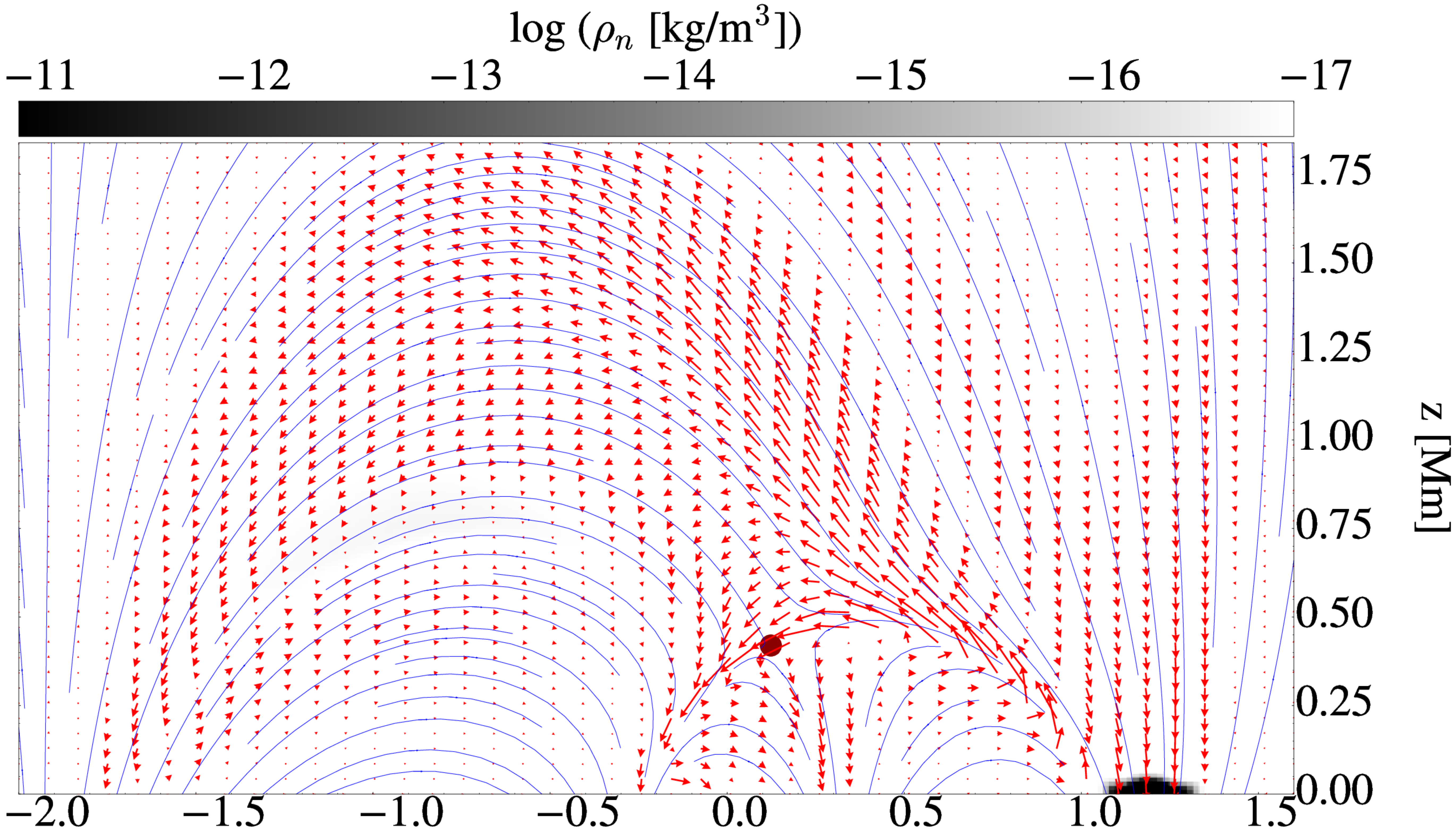}

\includegraphics[width=8cm]{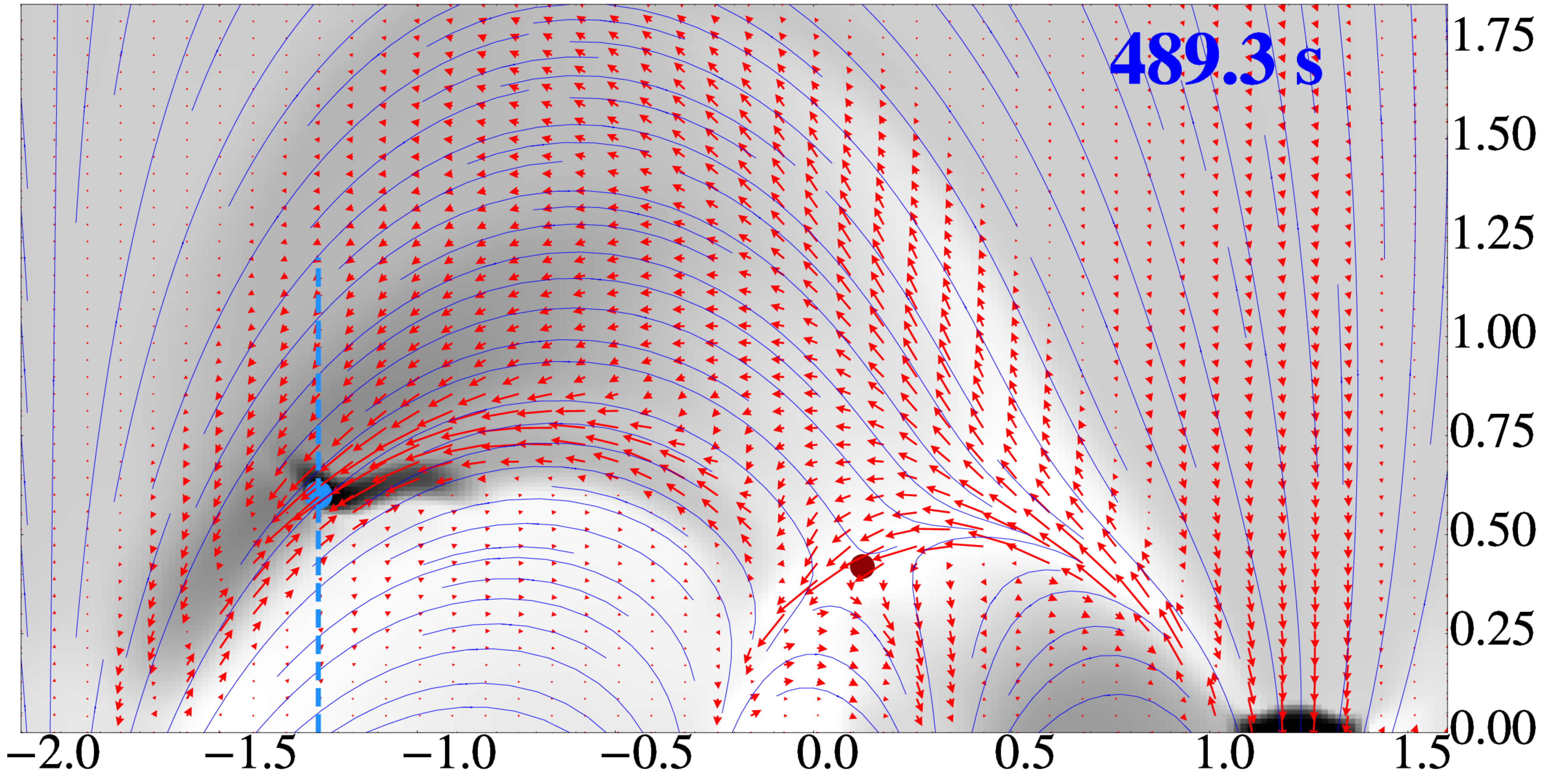}
\includegraphics[width=8cm]{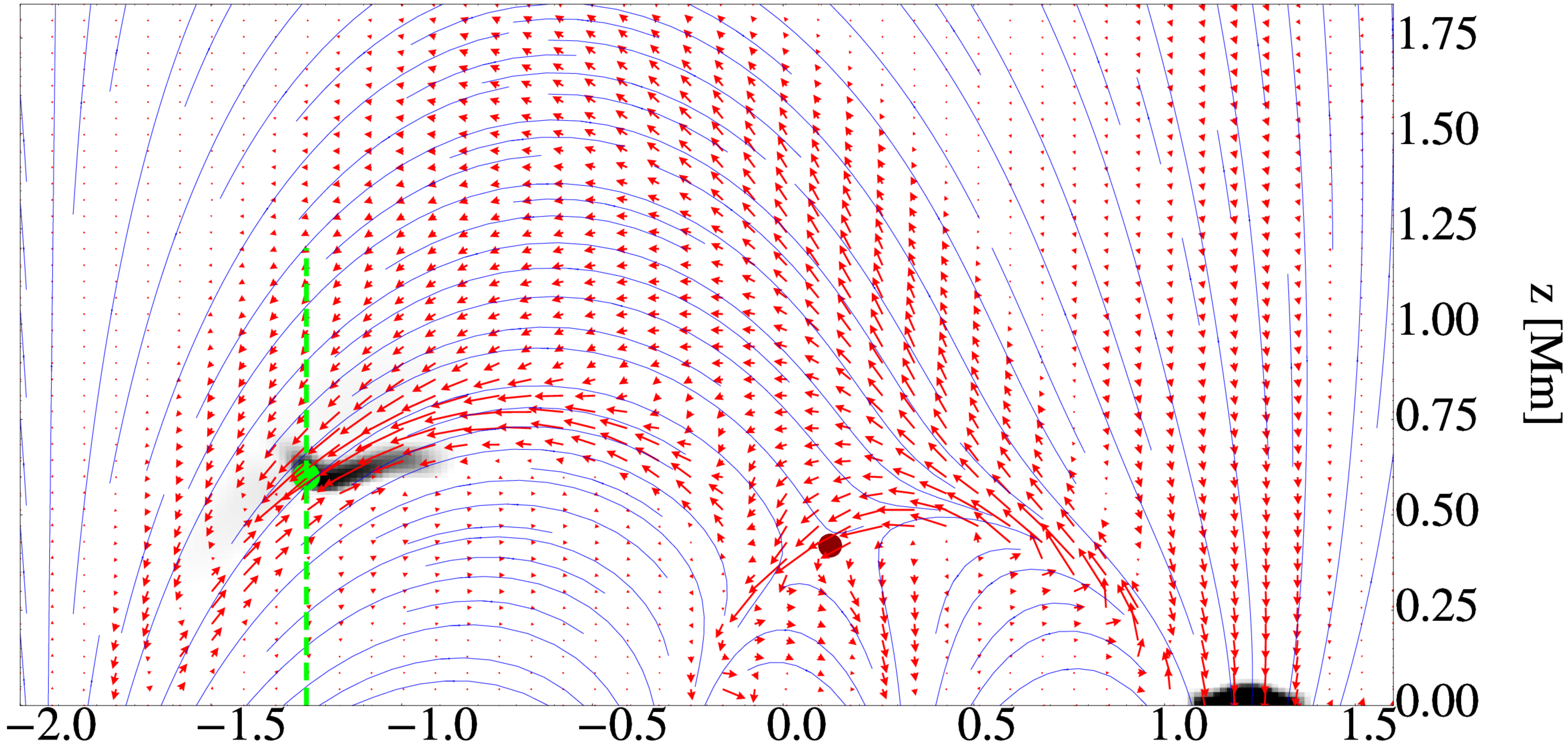}

\includegraphics[width=8cm]{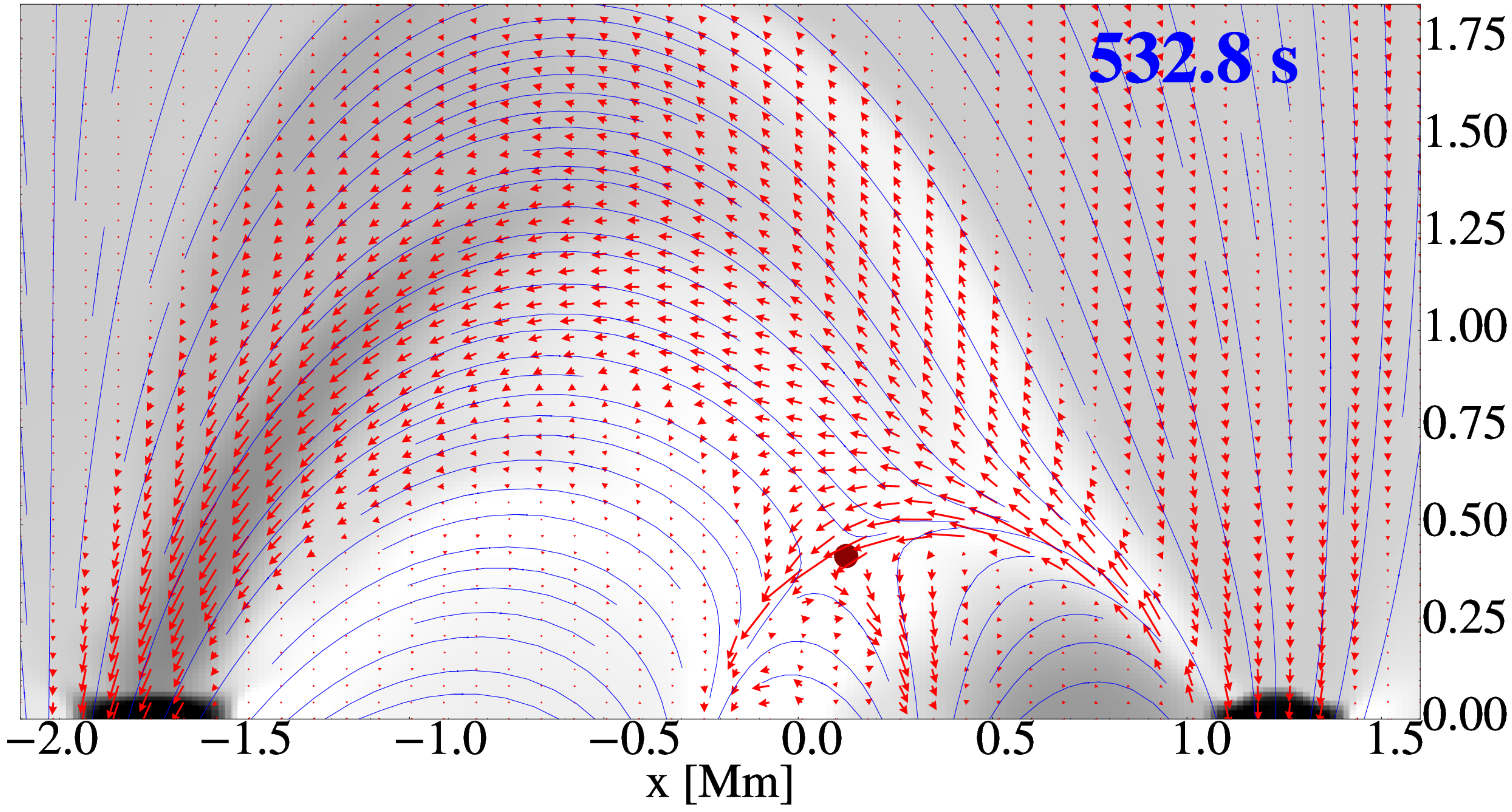}
\includegraphics[width=8cm]{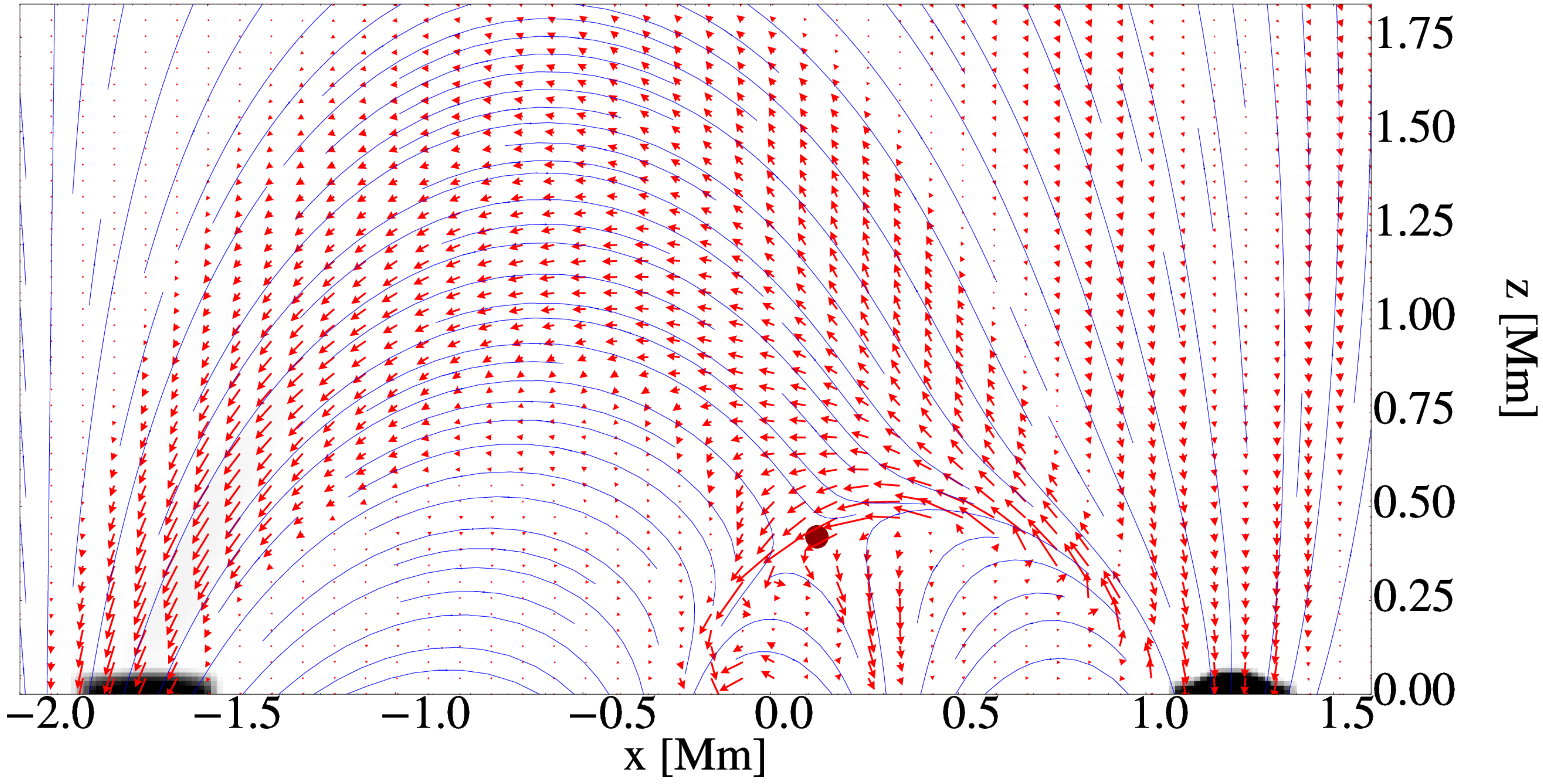}
}
\caption{Slice at $y=0$  for the times 
$t=442$~s (top row),  $t=489.3$~s (middle row) and $t=532.8$~s (bottom row), showing the formation and evolution of the rain blob.
 \textbf{Left:} Density of charges. The red arrows show the in-plane velocity of charges.
 \textbf{Right:} Density of neutrals.  The red arrows show the in-plane velocity of neutrals. 
The in-plane magnetic field lines are overplotted by blue lines in all the panels.
The perturbed null point, is shown by a red dot,
the other null point being outside the domain plotted in these panels.
At the time $t=489.3$~s, when the blob is formed fully (middle row)
the point with high density, situated at $x=-1.25$~Mm and $z=0.6$~Mm is marked in the charges and neutral density panels by a blue and green dot, respectively. 
This point is the same point marked in  Figure~\ref{fig:sli_x} and Figure~\ref{fig:rhoev} below for the panels on the second row, for a better visualization.
The line along the cut used in Figure \ref{fig:qline}, below is indicated by a  dashed line: blue in the charges density panel and green in the neutrals density panel.}
\label{fig:sli}
\end{figure*}
\begin{figure}
\FIG{
\includegraphics[width=8cm]{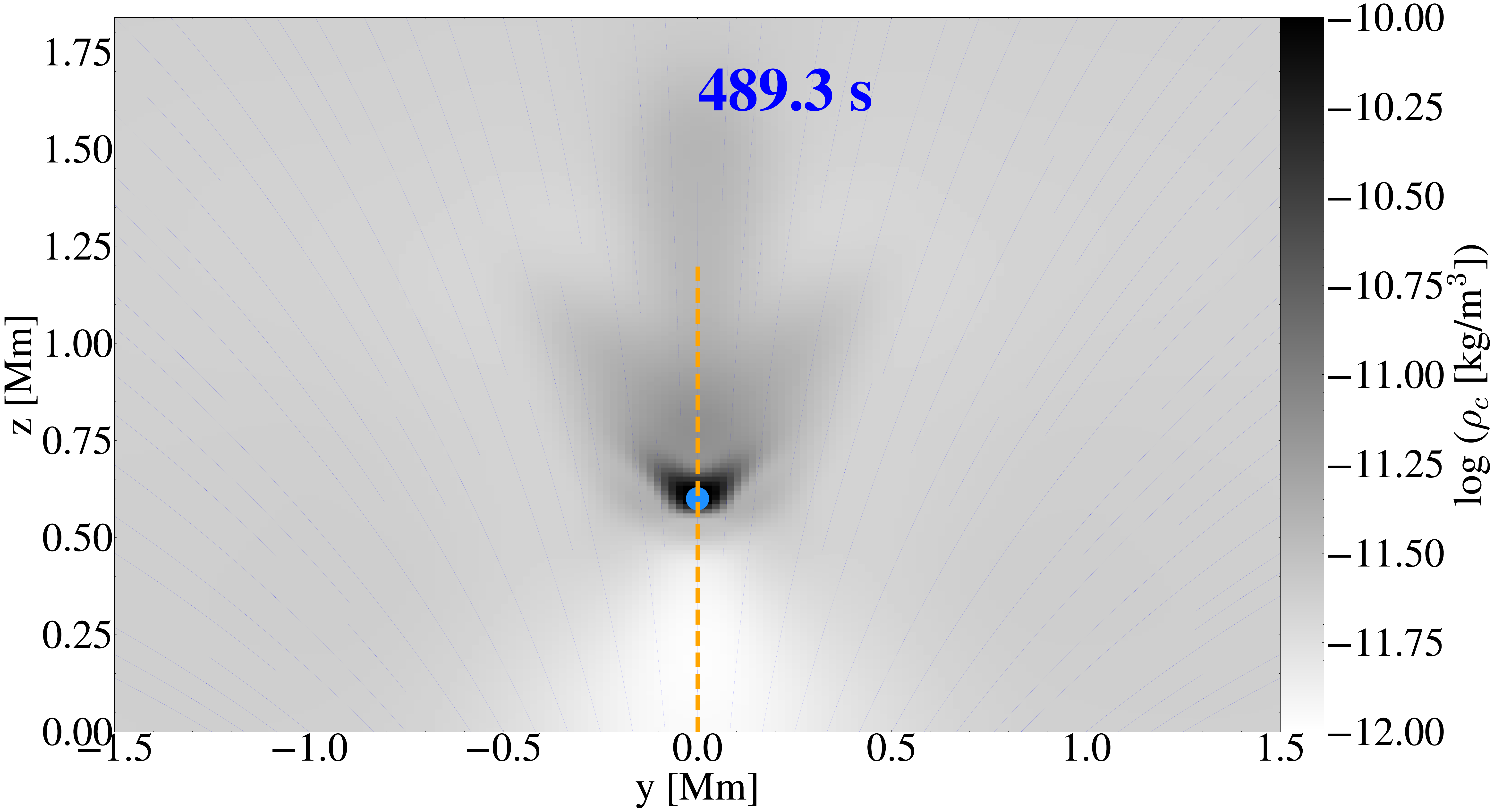}
}
\caption{Slice at $x=-1.25$~Mm for the time $t=489.3$~s showing the density of charges. 
The in-plane magnetic field lines are overplotted by blue lines. The point with high density, situated at $y=0$ and $z=0.6$~Mm is marked by a blue dot, being the same point
marked in Figure~\ref{fig:sli} above, and in
Figure~\ref{fig:rhoev} below for the panels on the second row, for a better visualization.
The line along the cut used in Figure \ref{fig:qline}, below is indicated by an orange dashed line.}
\label{fig:sli_x}
\end{figure}
\begin{figure}
\FIG{
\includegraphics[width=8cm]{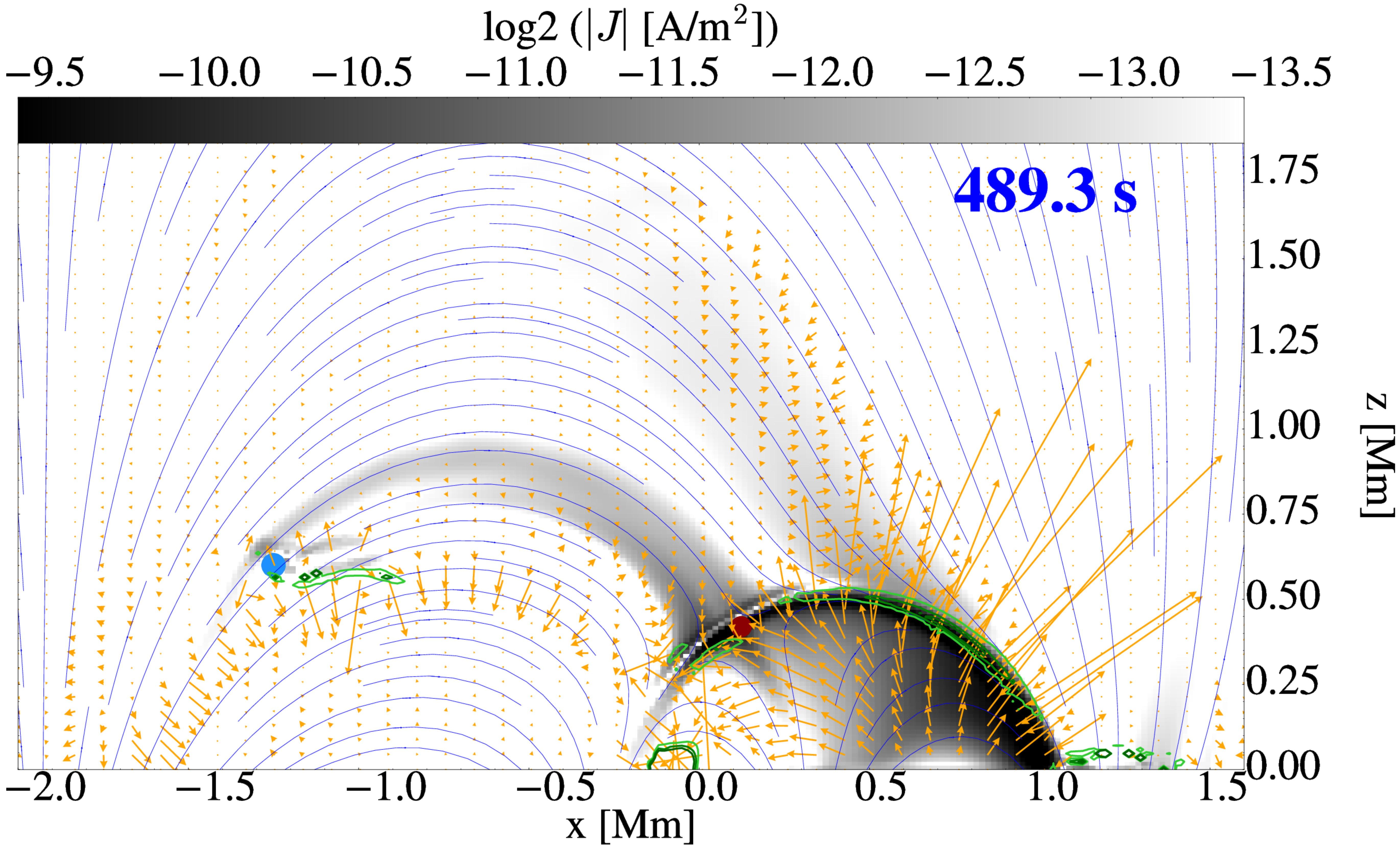}
}
\caption{Slice at $y=0$ for the time $t=489.3$~s of the magnitude of the current density. 
The quantity plotted is the base 2 logarithm of $|J|$, and the limits of the colorbar of -13.5 and -9.5
correspond to $|J|=8.6\times10^{-5}$~A/m$^2$ and $|J|=1.38\times10^{-3}$~A/m$^2$, respectively.
The in-plane magnetic field lines are overplotted by blue lines and the orange arrows indicate the in-plane decoupling velocity, the projection
in the $xz$ plane of the quantity $\vec{v_n}-\vec{v_c}$.
The perturbed null point is shown by a red dot. The isocontours of the decoupling in temperature $T_n-T_c$ are shown by green lines, indicating
three values of 10, 30 and 50 kK (darker green for larger value).
}
\label{fig:sli_j}
\end{figure}
\begin{figure*}
\FIG{
\includegraphics[width=8cm]{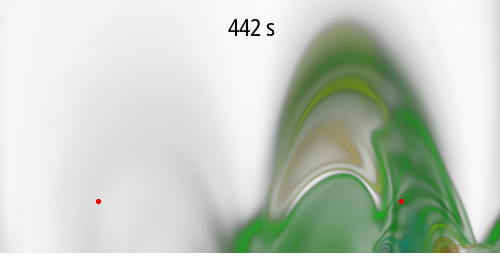}
\includegraphics[width=8cm]{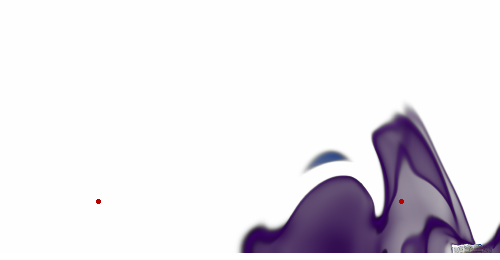}

\includegraphics[width=8cm]{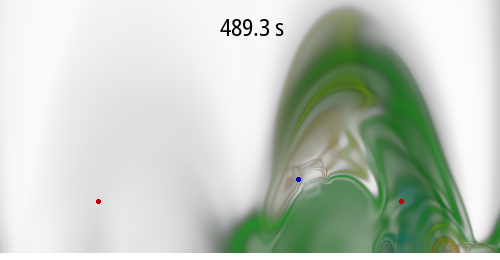}
\includegraphics[width=8cm]{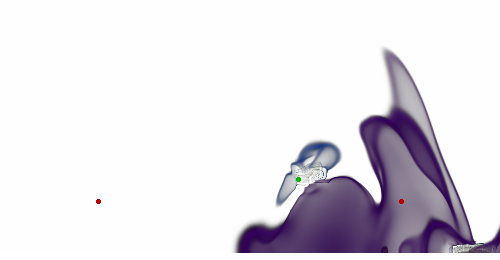}

\includegraphics[width=8cm]{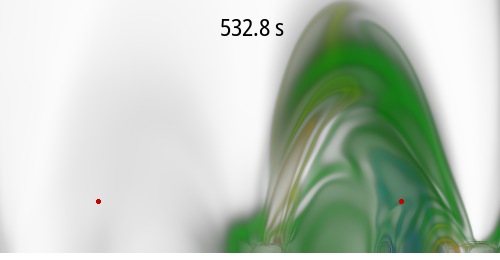}
\includegraphics[width=8cm]{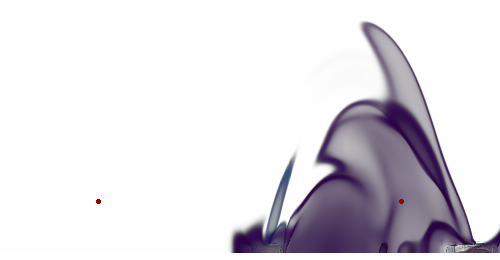}
}
\caption{Evolution of the density integrated along the $y$-direction around the time of the simulation when a 
condensation appear in the simulation. \textbf{Left:} Density of charges, \textbf{Right:} Density of neutrals. 
The magnetic null points are highlighted with small red circles. The 
point located at (-1.25,0,0.6)~Mm,  along the high density structure in the plane $y=0$ is marked for reference with green and blue, being the same points highlighted in 
Figures~\ref{fig:sli} and \ref{fig:sli_x} above. 
}
\label{fig:rhoev}
\end{figure*}
\begin{figure}
\FIG{
\includegraphics[width=8cm]{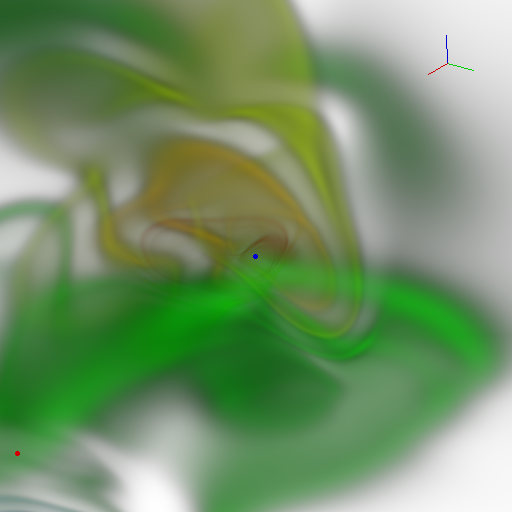}
}
\caption{3D integrated density of charges zoomed  around the condensation point at time $t=489.3$~s, shown from a different angle than in Figure~\ref{fig:rhoev}, as indicated by the colored axes.
The center of the condensation, 
located at (-1.25,0.0.6)~Mm is marked with a blue dot, and the perturbed null point is indicated by a red dot, consistently with Figures~\ref{fig:sli}, \ref{fig:sli_x} and \ref{fig:rhoev}. }
\label{fig:rhointzoom}
\end{figure}
\begin{figure*}
\FIG{
\includegraphics[width=16cm]{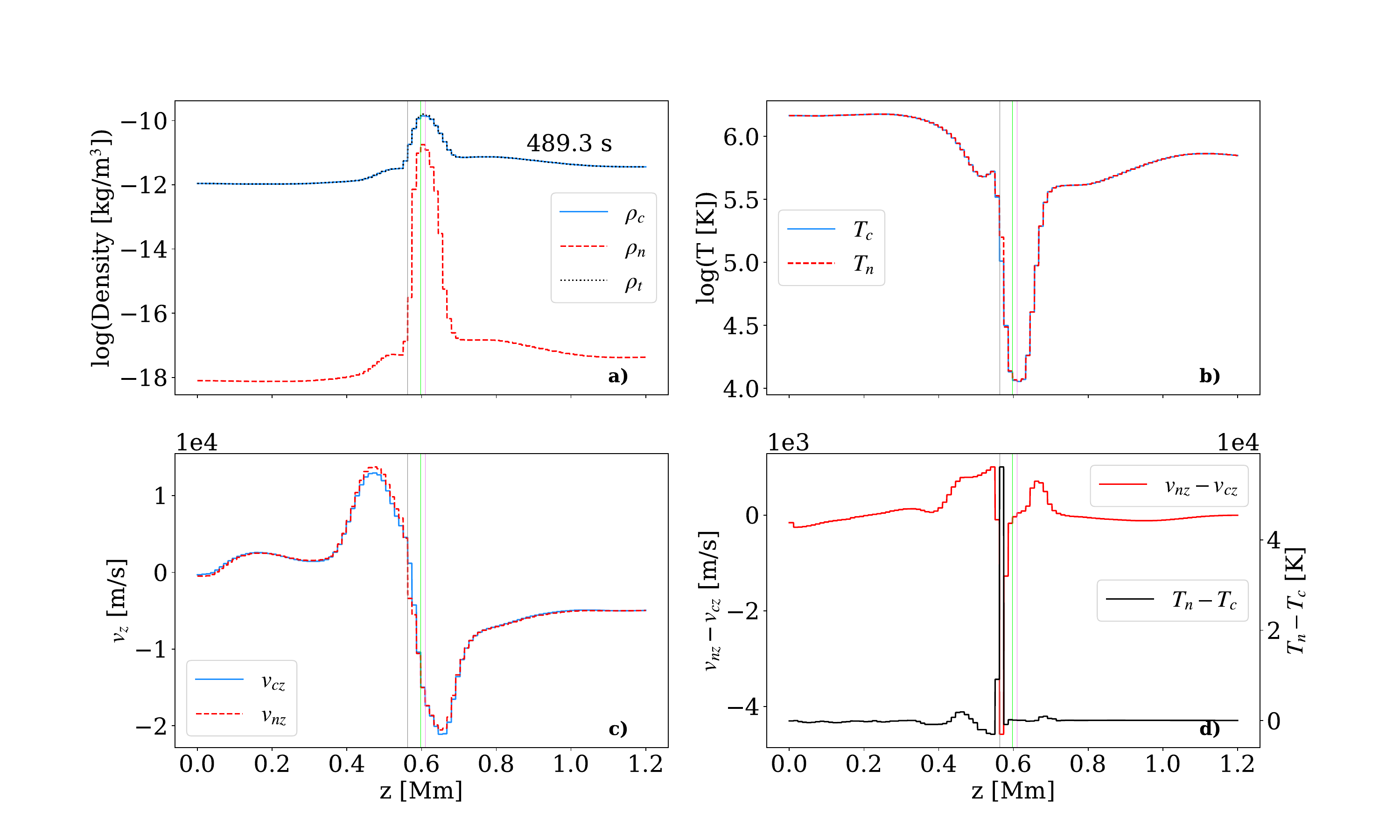}
}
\caption{Quantities along a vertical line cut at $x=-1.25$~Mm, $y=0$ for z between 0 and 1.2~Mm at time $t=489.3$~s.
\textbf{a)} Densities of charges (blue solid line), neutrals (red, dashed line) and total (black dotted line).
 \textbf{b)} Temperature of charges (blue solid line) and  neutrals (red, dashed line).  
 \textbf{c)}Vertical velocity of charges (blue solid line) and  neutrals (red dashed line).
 \textbf{d)} 
 Decoupling in velocity $v_{nz}-v_{cz}$ (red line on the left axis), and decoupling in temperature  $T_{n}-T_{c}$ (black line on the right axis). 
The three thin vertical reference lines correspond to: maximum density for all the three densities plotted: charges neutrals and total (green), maximum decoupling in both velocity and temperature (gray),
 and minimum temperature for charges and neutrals (violet).}
\label{fig:qline}
\end{figure*}

\subsection{Overall time evolution}
 
Figure~\ref{fig:tev} shows the time evolution of the current density at the perturbed null point, the maximum values in the whole domain of the density of charges and neutrals 
and the decoupling in velocity ($|\mathbf{v_n}-\mathbf{v_c}|$) between the two species. Although the time cadence in Figure~\ref{fig:tev} is relatively coarse, the trends are obvious: current density rises sharply at the null, and after several hundred seconds we witness the formation of coronal condensations in both charges and neutrals. Due to the bottom driving, the current density increases at the null point (panel a) shortly after the beginning of the simulation. 
The onset of thermal instability can be seen at around 300~s with a sudden increase of the maximum density of charges (panel b) and neutrals (panel d), when
the decoupling in velocity (panel c) also reaches the maximum value. The maximum decoupling in velocity shows a pattern similar to the current density at the null point, suggesting its relationship to the magnetic activity. 
The maximum density of neutrals shows a much steeper increase by seven orders of magnitude compared to almost two orders of magnitude that can be seen for the increase of the charge density.
The time scale related to the periodicity of the velocity driver (50~s, being P/2 used in Eq.~\ref{eq:vdr}) can be seen in the graphs of the evolution of the current density at the null point and the evolution of the velocity decoupling.  

\subsection{Coronal rain blob}

Since the magnetic null point is contained in the $y=0$ plane and the velocity driven at the bottom boundary only for the $x$-component has the largest magnitude at $y=0$ (and $x=x_{\rm N1}$),  we expect most dynamics due to the magnetic reconnection to happen in the $y=0$ plane.
Figure~\ref{fig:sli} shows a slice $y=0$ of the charges density (left panel) and neutral density (right panel) at three times, after the onset of the thermal instability:  $t=442$~s,  $t=489.3$~s and $t=532.8$~s, capturing  the formation and evolution of a condensation blob. 
The location where we encounter the maximum density for both neutrals and charges, located at $x=-1.25$~Mm and $z=0.6$~Mm is shown by a blue and green dot in the charges and neutral panels, respectively, for the time $t=489.3$~s, when the fully formed blob is traveling through the domain.

In the left panels we can observe field aligned structures due to the reconnection at the null point. 
The velocity shown by red arrows in the left panels indicate the movement of the  blob along the field lines from the reconnection point. 
At a subsequent time, $t=532.8$~s, the blob seems to have moved along the field lines and reaches the bottom boundary at $x\approx-1.75$~Mm, where its further movement is slowed down by the zero velocity bottom boundary conditions, similarly to a previously formed blob, which can be seen at $x\approx1.25$~Mm in all the panels. The two blobs have similar sizes and shapes.

Previous to the blob formation, at time $t=442$~s, we can observe larger density above the location of the blob, indicating that most of the condensed material will come from above.  
The reconnection jets travel along the current sheets formed in the vicinity of the null point, 
and there is a spatial correlation between the charged density and the current density, shown in Figure~\ref{fig:sli_j} below for the time $t=489.3$~s.
The density of charges seems to be larger along the spine, 
however, the recombination of charges show a depletion on the path traveled by the previously formed blob along the dome surface.
We can also observe the spine becoming more inclined with time, consistent to the collapse of the null point during the shearing driving \citep{priest1,pontin1}.

The spatial contrast in the density of neutrals, much larger than that in the charged density, suggested by the limits of the colorbars, indicates the important role of the recombination.
Because of the  recombination, the neutrals  slow down and the charges speed up. We will see how this effect, and the fact that the magnetic forces act on the charges only, affect the decoupling at different scales.

To infer the 3D shape of the blob we look at a $x$-slice when it is formed.
Figure~\ref{fig:sli_x} shows a slice of the charged density at $x=-1.25$ at time $t=489.3$~s, indicating
that the condensation has sizes of the same order of magnitude of several hundreds kilometers for all the dimensions, consistent 
with the observations \citep{obsCR1,antolin2023}. 
The condensations align perpendicularly to the field lines, especially when the magnetic field is strong \citep{ti2,ti2f}, 
shrinking in time in the direction parallel to the field lines. 
The blob is further deformed and it appears partially along the field lines, as seen in the $y=0$ slice because of the high velocity reconnection outflows in this plane, 
indicated by large red arrows, corresponding to velocities of around 50-60~km/s.

The linear theory shows that the thermal instability growth rate is weakly sensitive to the wavelength of the perturbation \citep{field}, 
therefore we expect the blobs to appear in the inverse order of their 
initial perturbation amplitude.
The condensation blobs appear in the nonlinear stage of the thermal instability, as a result of the growth of the density perturbations created by reconnection.
Therefore the sizes of the initial perturbation are related to the 
sizes of the reconnection jets, which in turn are related to the width of the current sheets.
Figure \ref{fig:sli_j} shows the magnitude of the current density, in a base 2 logarithm, a decrease by 1 meaning that the current density magnitude dropped to half, therefore
making easier the estimation of the full width at half maximum (FWHM), an indicator for the current sheet width.
The current sheet contains part of the fan and spine,
being stronger at the dome boundary along the separatrix surface than along the spine, 
similarly to the current density structures observed in Figure~9 in \cite{Pontin_2013}.
The current sheet width seems to be similar to the size of the condensed structures in the direction perpendicular to the field lines.
While the thermal instability evolves the blobs shrink in the direction parallel to the field lines.

We overplot the in-plane decoupling in velocity ($\mathbf{v_n}-\mathbf{v_c}$ projected in the $xz$ plane) with orange lines and the isocontours of the decoupling in temperature
($T_n -T_c$) by green lines, showing three values of 10, 30 and 50 kK with progressively darker colors.
The decoupling in velocity shows greater values around the blob in the bottom part of the domain, where the magnetic field is stronger. 
The neutral particles go inside the condensed structure across the magnetic field lines, unaffected by the Lorentz force. 
At the same time, we observe the highest decoupling in velocity 
 above the   
previously formed blob ($x\approx1.25$~Mm), on its path along the field lines towards the bottom.  
The large velocity decoupling, correlated spatially with a high current density (associated with large Lorentz force) 
and with the depletion of the charged density, seen in the corresponding left panel of Figure~\ref{fig:sli},  
suggests that the decoupling is related to the recombination of the charges and the slippage of neutrals across the field lines.
The decoupling in velocity produces frictional heating, which increases the temperature of the neutrals more than that of the charges, because of their lower density,
showing a difference of more than 50~kK, at the location where the decoupling in velocity is maximum.

A better view of the 3D density structures can be seen in Figure~\ref{fig:rhoev}, which shows the density of charges (left panels) and neutrals (right panels), integrated along the $y$-direction for the three times shown before in Figure~\ref{fig:sli}, around the time when the condensation formed in both charges and neutrals.
The two null points are indicated by two red dots, and the location where we find the maximum density in the slice $y=0$,  for both neutrals and charges at time $t=489.3$~s, 
situated at $x=-1.25$~Mm and $z=0.6$~Mm, is shown by a blue and green dot in the charges and neutral panels, respectively, similarly to Figures~\ref{fig:sli} and \ref{fig:sli_x} above. 
The high density rain blob surrounds this maximal density location.
The charged density structures can 
be seen at the second null point as well, but much less dense and with no condensations.
The difference in the contrast of density between charges and neutrals can also be seen spatially. The integrated density of neutrals seems to have a much higher contrast, with sharp structures, compared to the charged density. No structure in the neutral density can be visually seen around the second null point. The long and thin 3D  structures  in the charged density, from which the blob condenses can be better seen in a zoom-in around the high density point, shown in Figure~\ref{fig:rhointzoom}. They are consistent with their 2D projection, the  flows along the spine, seen in Figure~\ref{fig:sli}.

\subsection{Scales of ion-neutral decoupling}

In order to quantify better the  density contrast and the two-fluid effects, we look at a cut along the line $x=-1.25$~Mm and $y=0$ for $0 \le z \le 1.2$~Mm, containing the point with the highest density located at $z=0.6$~Mm, already mentioned and indicated in the previous figures. Panel a) of Figure~\ref{fig:qline}
shows the density of charges, neutrals and the total density along this line.
We can observe that the contrast in neutral density is seven orders of magnitude indicating a major increase compared to its original coronal and negligible values. It is also more pronounced when compared to the contrast in the charges density, of two orders of magnitude. The latter is the typical density contrast between corona and the interior regions of prominences and coronal rain. Still, throughout the blob we have always at least one order of magnitude higher plasma than neutral density. Therefore, 
the neutral density remains small compared to the charges density so that the total density contrast is that of the charges. 

The temperatures of charges and neutrals have a similar profile (panel b), with no visible difference, 
so plasma-neutral thermal coupling is near perfect and the blob represents a hundredfold colder material than the corona, the typical observed contrast. 
The minimum peak in the temperatures (the vertical violet line) is slightly shifted up in the vertical direction from the maximum peak in the densities (the vertical green line).
The distance between these two peaks is around 12~km, which is similar to the value of the collisional mean free path between ions and neutrals \citep[as from Eq.~(37) in][where we used the charges fast magneto-acoustic speed]{paper2f}, calculated at the center of the blob, suggesting that 
this misalignment could be explained by the two-fluid effects. 
Outside the blob the density of neutrals is very small,  therefore the collisional mean free path between ions and neutrals is very large (of the order $10^9$~km), orders of magnitude larger than the simulation domain.
The larger density of charges makes the collisional mean free path between neutrals and ions \citep[Eq.~(37) in][where we used the neutrals sound speed]{paper2f} smaller
than that between ions and neutrals, being about 80~km outside the blob and reaching a value smaller than 1~km in the center of the condensation.

When the condensation forms, the charged and neutral particles accumulate inside the blob with velocities of about 10~km/s, which can be seen in the velocity profile, shown in panel c).  The whole structure moves downwards, with the charges moving slightly faster, everywhere, except for a small layer in the center of the structure. The downward velocity of about 10~km/s is similar to observations \citep{obsCR1,antolin2023}.

The decoupling in velocity and temperature, shown in panel d) have the same peak in magnitude (the vertical gray line), located towards the bottom, compared to the peak in the densities. 
The negative peak in the decoupling in velocity appears due to the movement of the neutrals across the field lines, when both neutrals and charges go towards the center of the moving blob, mostly from above.
However, we can observe two smaller positive peaks in the decoupling at the sides, occurring at larger scales, 
when the neutrals are slowed by the recombination in the downward moving blob.
The frictional heating, proportional to the square of the decoupling in velocity, is much more efficient to heat the neutrals because of their lower density compared to that of the charges.
This difference becomes reflected in the decoupling in temperature with the maximum located at the same point as the maximum decoupling in velocity, 
and the neutrals having larger temperature than the charges. 
The maximum decoupling in both velocity and temperatures are larger, over 10\% of the values of velocity and temperature, respectively: 
about 1~km/second in velocity decoupling and few 10000~K in temperature. 
Because of the increase in temperature due to the frictional heating, the minimum temperature peak is shifted slightly towards the upper part.
The radiative cooling also becomes more efficient for slightly higher temperatures, because of the dependence of the cooling curve on the temperature.

\section{Conclusions}
\label{sec:concl}

In this paper we demonstrated for the first time in a full two-fluid setting, the reconnection induced formation of coronal rain, in a 3D realistic setup  of spine-fan reconnection. 
The shearing motions at the boundary will produce spine-fan reconnection, the spine collapses towards the fan, and a strong current sheet forms in the vicinity of the null point, 
containing parts of both fan and the spine. Even though the magnetic field configuration does not represent a ``proper radial null point'', usually considered in the analytical 
studies \citep{priest1,pontin1}, and the spine is inclined with respect to the plane tangential to the fan, the evolution is similar. The spine becomes
progressively more inclined, similar to a collapse of the radial null point. 

The thermal instability linear growth rate is not very sensitive to the scale size \citep{field,ti2f},
the shape, size and amplitude of the perturbations leading to the rain blob being determined by the reconnection process.
The  perturbations created by this reconnection scenario make the condensations appear earlier than in other studies of coronal rain \citep{xiaohong_cr}, 
and we witness blobs formed away from the perturbed null region after several minutes in the process. 
The blob formation time is inversely proportional to the amplitude of the perturbation, which in turn depends on 
many quantities, such as the magnetic field, amplitude of the driving, resistivity and plasma beta.
The growth of the thermal instability depends mainly on the cooling curve, which is a function of the temperature.
The blob has a shape, size and speed consistent with the observations. 
Another blob, formed previously, has similar size and shape.
We did not have gravity in our simulation, therefore the speed of the blobs in our simulation depends mostly on the Lorentz force. The previous blob which traveled along
the fan surface might have had larger velocity because of the larger current density. 
The resemblance to observations is due to the fact that coronal rain is usually observed as a result of 
3D null point reconnection \citep{Mason_2019}, and we used values for density and temperature, magnetic field 
configuration and driver velocity similar to observations of the solar corona.

The current density evolution at the null point is sensitive to the boundary driving and higher currents are generally  correlated to higher decoupling in velocity. 
A steep increase in time in the maximum density of neutrals and charges occurs when the structure condenses to smaller scales, and the decoupling in velocity reaches its maximum. 
The perturbations around the secondary null point, which is not directly driven, are small and do not form condensations during the simulation time. 
The temperature inside the condensing blobs drops and triggers the recombination. 
The density of the neutrals throughout the blob remains smaller than that of the charges, and the overall density contrast is that of the charges which 
is about two orders of magnitude, consistent with the observations. 
However, the contrast in the neutral density is almost five orders of magnitude  larger than that in charges. 
Even if the overall density contrast is not modified by the presence of neutrals, the decoupling in velocity produces frictional heating, which can be significant, 
being proportional to the square of difference in velocity. The decoupling between neutrals and charges can be observed at two different scales.
Recombination slows the neutrals down and accelerates the charges, seen in larger downward velocity of charges than of neutrals for the overall moving blob, 
producing decoupling at larger scales.
However, in a very small layer inside the blob,  the neutrals entering the structure mainly from above (because the blob is moving downwards), 
move faster across the field lines, compared to the charges, and the
decoupling at these smaller scales has opposite sign to that at larger scales.

The two fluid effects create small scale structures due to very localized frictional heating, modifying further the temperature profile and consequent radiative cooling, taking into account that the cooling curves are very sensitive to the temperature.
The two-fluid effects become more important inside the condensed blobs where the neutral density increases fast due to the recombination and  
there is large decoupling in velocity and  temperature.

The width of the small scale structures created by the two-fluid effects due to elastic collisions is comparable to the mean free path between ions and neutrals, 
which is resolved in our simulation. However,  the mean free path between neutrals and ions is not resolved at the center of the condensation,
underestimating the collisional effect of the charges on the neutrals evolution.
A better resolution would improve the understanding of this effect, however a resolution of 1~km in a 3D setup is computationally challenging.
We focused on the formation and evolution of only one blob, mentioning another previously formed blob. 
We expect two-fluid effects to be general for all blobs in this simulation, 
and simulations of much longer duration can address statistical properties of blob sizes, typical characteristics, and overall dynamics.

The simulation does not include gravity, mainly because it is difficult to construct a stratified atmosphere in both mechanical and ionization/recombination 
equilibrium for both neutrals and charges, which should be overcome in future work. In such more realistic, stratified settings, it will be relevant to look at rain cycles, 
while this work serves as a proof-of-concept that thermal instability can occur in a spine-fan setup, accounting for relevant plasma-neutral effects.

\begin{acknowledgements}
This work was supported by the FWO grant 1232122N and European Union (ERC, DynaMIT, 101086985). RK acknowledges the KU Leuven C1 project C16/24/010 UnderRadioSun and the FWO project G0B9923N Helioskill.
The resources and services used in this work were provided by the VSC (Flemish Supercomputer Center), funded by the Research Foundation - Flanders (FWO) and the Flemish Government.
\end{acknowledgements}
\bibliographystyle{aa}

\end{document}